\documentclass[aps,prb,twocolumn,floatfix,showpacs,citeautoscript,longbibliography,hyperlinks]{revtex4-1}

\usepackage{amsmath}
\usepackage{epstopdf}
\usepackage{amsmath}
\usepackage{amsthm}
\usepackage{amssymb}
\usepackage{amsfonts}
\usepackage[colorlinks=true,citecolor=blue]{hyperref}
\usepackage{graphicx,graphics}
\usepackage{physics}
\usepackage{float}
\usepackage[caption = false]{subfig}

\begin{document}

\title{Optimal entanglement witness for Cooper pair splitters}
\author{Minh Tam}
\author{Christian Flindt}
\author{Fredrik Brange}
\affiliation{Department of Applied Physics, Aalto University, 00076 Aalto, Finland}

\begin{abstract}
The generation of spin-entangled electrons is an important prerequisite for future solid-state quantum technologies. Cooper pairs in a superconductor can be split into separate electrons in a spin-singlet state, however, detecting their entanglement remains an open experimental challenge. Proposals to detect the entanglement by violating a Bell inequality typically require a large number of current cross-correlation measurements, and not all entangled states can be detected in this way. Here, we instead formulate an entanglement witness that can detect the spin-entanglement using only three cross-correlation measurements of the currents in the outputs of a Cooper pair splitter. We identify the optimal measurement settings for witnessing the  entanglement, and we illustrate the use of our entanglement witness with a realistic model of a Cooper pair splitter for which we evaluate the cross-correlations of the output currents. Specifically, we find that the entanglement of the spins can be detected even with a moderate level of decoherence. Our work thereby paves the way for an experimental detection of the entanglement produced by Cooper pair splitters.
\end{abstract}

\maketitle

\section{Introduction}

Cooper pair splitters are promising devices for generating spin entanglement between separated electrons in solid-state systems. A Cooper pair splitter takes advantage of the entangled electrons that naturally form as Cooper pairs inside a superconductor \cite{Lesovik2001,Recher2001}. The pairs can be extracted from the superconductor by coupling it to quantum dots with strong Coulomb interactions, which force the pairs to be split between them. Several experiments have demonstrated the extraction and splitting of Cooper pairs in different solid-state architectures \cite{PhysRevLett.93.197003,PhysRevLett.95.027002,Hofstetter:Cooper,PhysRevLett.104.026801,Wei2010,PhysRevLett.107.136801,PhysRevLett.109.157002,Herrmann:Spectroscopy,Das2012,PhysRevB.90.235412,PhysRevLett.114.096602,PhysRevLett.115.227003,Borzenets:High,Bruhat2018,tan2020,Ranni2021} involving quantum dots~\cite{Hofstetter:Cooper,PhysRevLett.107.136801,Herrmann:Spectroscopy}, carbon nanotubes~\cite{PhysRevLett.104.026801}, or graphene nanostructures \cite{PhysRevLett.115.227003,Borzenets:High,tan2020,PhysRevLett.126.147701}. The splitting of Cooper pairs has been confirmed through measurements of the non-local conductance or the low-frequency noise \cite{Wei2010,Das2012}, and very recently with single-electron detectors \cite{Ranni2021}. Still, a direct detection of the spin entanglement remains an outstanding experimental challenge.

Detection schemes based on Bell inequalities\cite{doi:10.1143/JPSJ.70.1210,PhysRevB.66.161320,PhysRevLett.91.157002,PhysRevLett.91.147901,PhysRevLett.92.026805,Braunecker2013,PhysRevLett.114.176803,PhysRevB.96.064520} or full quantum state tomography\cite{PhysRevB.73.041305} have been proposed and formulated in terms of current cross-correlations measurements.\cite{Blanter20001} However, these schemes typically rely on a large number of measurements, for instance, each of the four correlators in a standard Bell inequality requires four different cross-correlation measurements, such that a total of 16 different measurements are needed.\cite{doi:10.1143/JPSJ.70.1210,PhysRevB.66.161320,PhysRevLett.91.157002,PhysRevLett.91.147901,PhysRevLett.92.026805,PhysRevLett.114.176803} Moreover, Bell inequalities are designed to test the concept of local realism, and some entangled states cannot be detected in this way.\cite{PhysRevA.40.4277} Thus, to provide an alternative path towards the detection of entanglement, the use of entanglement witnesses \cite{Terhal2000319,PhysRevA.66.062305,PhysRevLett.94.060501,Guhne20091} has been proposed \cite{PhysRevB.75.165327,PhysRevB.89.125404,Baltan_s_2015,Brange2017}. An entanglement witness is an observable whose expectation value for one entangled state is different from the expectation values of all separable states \cite{Terhal2000319}. Earlier work has found that certain spin-entangled states can be witnessed using only two cross-correlation measurements.\cite{Brange2017} However, it is also known that the singlet state, which is maximally entangled, surprisingly cannot be detected in this way.\cite{PhysRevA.81.052339,Brange2017} 

\begin{figure}[b]
    \begin{subfloat}
        \centering
        \includegraphics[width=0.47\textwidth]{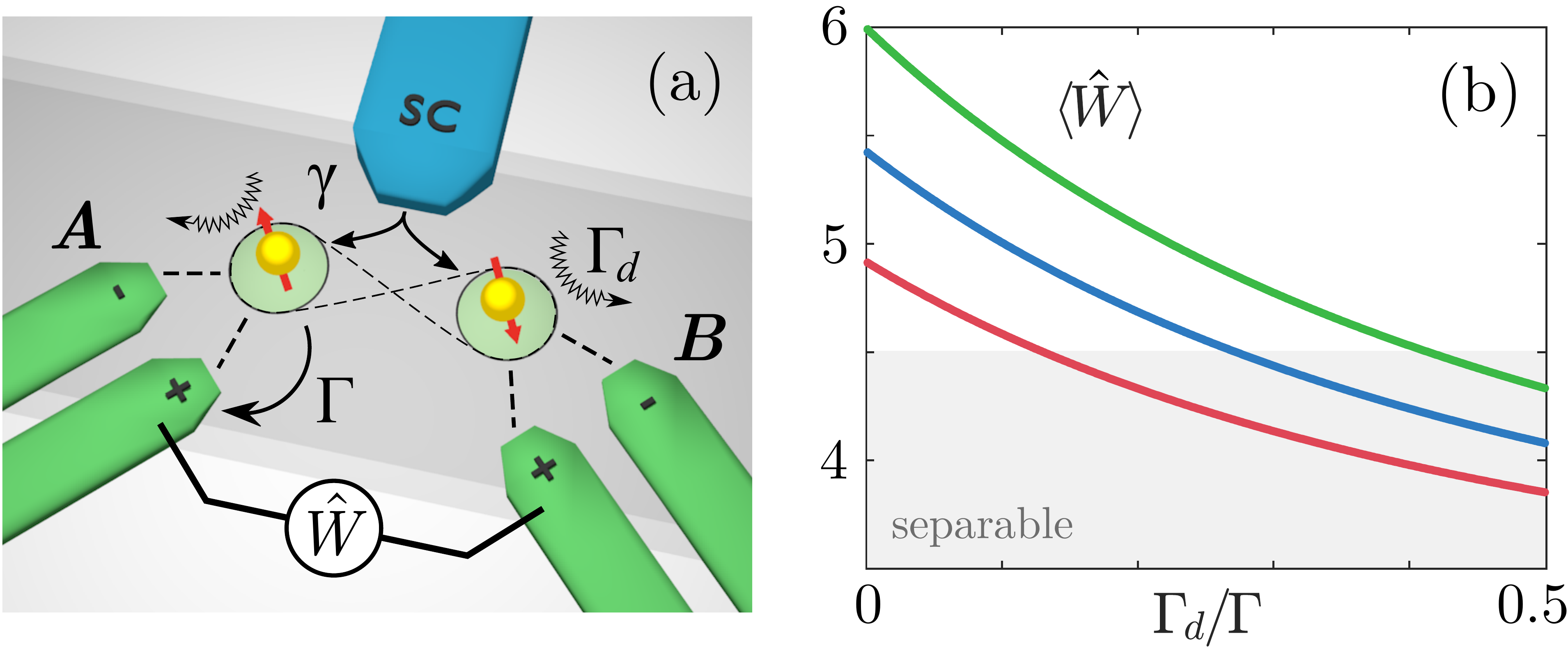}
        \captionsetup{justification=justified,singlelinecheck=false}
        \caption{(a) Schematics of a Cooper pair splitter consisting of a superconductor (blue) that emits electrons into a double quantum dot (light green) at the rate $\gamma$. The energy levels $\varepsilon_i$ of the dots $i=A,B$ are detuned so that elastic cotunneling with amplitude $\kappa$ is suppressed, $|\varepsilon_A-\varepsilon_B|\gg \kappa$. The decoherence rate inside the quantum dots is denoted as $\Gamma_d$. The dots are coupled at the rate $\Gamma$ to two pairs of ferromagnetic leads (green), which act as spin-sensitive detectors, $A$ and $B$. The entanglement witness $\hat W$ is based on three different current cross-correlation measurements between the output currents in the leads denoted by $A+$ and $B+$. (b) The expectation value of the entanglement witness with optimal detector settings as a function of the ratio $\Gamma_d/\Gamma$. Here $\gamma = \kappa$, $\Gamma = 100\gamma$, $\varepsilon_A=-\varepsilon_B=5\gamma$, and the (unknown) detector efficiencies are $\zeta_A=\zeta_B=1$ (green), $\zeta_A=\zeta_B=0.9$ (blue) and $\zeta_A=\zeta_B=0.8$ (red). The gray-shaded area indicates the expectation values that can be obtained from separable (non-entangled) states.}
        \label{Fig1}
    \end{subfloat}
\end{figure}

In this work, we consider the Cooper pair splitter illustrated in Fig.~\ref{Fig1}(a) and formulate a witness that can detect the entanglement of split Cooper pairs using merely three current cross-correlation measurements. We find that the optimal settings of the detector systems are to position the polarization vectors for each of the three measurements radially symmetrically in a plane. Importantly, for spin read-out with ferromagnetic leads, this means that it is not necessary to rotate the magnetic polarization in all three dimensions. We illustrate our entanglement witness with a model of a Cooper pair splitter, which allows us to evaluate the current cross-correlations and investigate the effects of decoherence on the detection of entanglement. In Fig.~\ref{Fig1}(b), we show the entanglement witness as a function of the decoherence rate over the coupling to the leads, and we see that the entanglement can be detected even with moderate levels of decoherence compared to the tunneling rates. These result were obtained for three unknown detector efficiencies, and as we will see in the following, the detector margin improves further, if the detector efficiencies are experimentally known. We will also discuss the experimental perspectives of detecting the entanglement generated by a Cooper pair  splitter using our entanglement witness. As such, our work provides a feasible way towards the experimental detection of the entanglement produced by Cooper pair splitters.

The rest of the paper is organized as follows. In Sec.~\ref{Cooper pair splitter}, we introduce the  model of a Cooper pair splitter including the detector systems and decoherence. In Sec.~\ref{Noise and correlations}, we evaluate the average currents and their cross-correlations using methods from full counting statistics. Based on the current cross-correlations, we formulate in Sec.~\ref{Entanglement witness} an entanglement witness that can detect the entanglement of pairs of electrons in a singlet state. In Sec.~\ref{Detection of the singlet state}, we go on to investigate the properties of the witness and identify the optimal settings that maximize the detection margin. In Secs.~\ref{Detection of less entangled pure states} and~\ref{Detection of decohered states}, we consider the entanglement detection of other pure states than the singlet state as well as of mixed states, respectively. In Sec.~\ref{exp_pers}, we return to the model of a Cooper pair splitter and discuss the experimental perspectives of detecting the entanglement produced by Cooper pair splitters. Here, we find that the entanglement can be detected even with a moderate level of decoherence in the system. Finally, in Sec.~\ref{Conclusion and outlook}, we summarize our work.

\section{Cooper pair splitter}
\label{Cooper pair splitter}
We consider the Cooper pair splitter in Fig.~\ref{Fig1}(a), consisting of two single-level quantum dots in close proximity to a conventional spin-singlet $s$-wave superconductor. The superconductor acts as a source of Cooper pairs, which injects pairs of electrons in a singlet state into the quantum dots. The pairs are split into different dots due to strong on-site Coulomb interactions that prevent each quantum dot from being occupied by more than one electron at a time. Each quantum dot is also tunnel-coupled to two ferromagnetic leads with opposite polarizations, serving as drains for the split Cooper pairs. The leads act as spin-sensitive detectors, with the probabilities for an electron to tunnel into each lead set by the spin projection onto the polarization vectors. Electrons in the quantum dots may also interact with the environment, leading to decoherence.

For a large superconducting gap, the transport between the superconductor and the quantum dots is dominated by Cooper pair splitting and elastic cotunneling, whose (real) amplitudes we denote by $\gamma$ and $\kappa$, respectively. The coherent dynamics of the quantum dots is then described by the effective Hamiltonian\cite{Sauret:Quantum,Eldridge:Superconducting,Hiltscher2011,Walldorf2020}
\begin{equation}
\hat{H}=\!\sum_{\ell\sigma}\varepsilon_\ell\hat{d}_{\ell\sigma}^\dagger \hat{d}_{\ell\sigma}^{\phantom\dagger}-\bigg(\gamma\hat d_S^\dagger \!+\!\sum_\sigma\!\kappa \hat{d}_{A\sigma}^\dagger \hat{d}_{B\sigma}^{\phantom\dagger}+\text{H.c.}\!\bigg),
\end{equation}
where $\varepsilon_\ell$ is the energy level of dot $\ell = A,B$, $\hat d_{\ell\sigma}^\dagger$ is the creation operator for an electron with spin $\sigma = \uparrow,\downarrow$, and
\begin{equation}
\hat d_S^\dagger \equiv (\hat{d}_{A\downarrow}^\dagger \hat{d}_{B\uparrow}^\dagger-\hat{d}_{A\uparrow}^\dagger \hat{d}_{B\downarrow}^\dagger)/\sqrt{2}
\end{equation}
is the creation operator for the two-electron spin-singlet state. The detuning $|\varepsilon_A-\varepsilon_B|\gg\kappa$ is taken so large that elastic cotunneling is strongly suppressed, while $\varepsilon_A+\varepsilon_B=0$ so that Cooper pair splitting is on resonance.

With a large bias driving electrons unidirectionally out of the dots to the leads, the full time-evolution of the quantum dots, with the ferromagnetic leads and decoherence included, is described by the Lindblad equation \cite{breuer2007theory,Flindt2005, RevModPhys.81.1665,Malkoc2014}
\begin{equation}
\frac{d}{dt}\hat \rho_t=\mathcal{L}\hat \rho_t=-\frac{i}{\hbar}[\hat H,\hat \rho_t]+\mathcal{D}_\mathrm{fer}\hat \rho_t+\mathcal{D}_\mathrm{dec}\hat \rho_t,
\label{Lindblad equation}
\end{equation}
where $\mathcal{L}$ is the Liouvillian, $\hat{\rho}_t$ is the (time-dependent) density matrix of the quantum dots, $\mathcal{D}_\mathrm{fer}$ is a dissipator describing the ferromagnetic leads, and $\mathcal{D}_\mathrm{dec}$ is a (set of) dissipators describing environment-induced decoherence.

The dissipator for the ferromagnetic leads reads\cite{Malkoc2014}
\begin{equation}
\mathcal{D}_\mathrm{fer}\hat \rho_t=\Gamma\sum_{\ell \sigma}\left(\sum_{m \sigma'}\mathcal{J}_{\ell m}^{\sigma \sigma'}\hat \rho_t-\frac{1}{2}\{\hat \rho_t,\hat d_{\ell\sigma}^\dagger \hat d_{\ell\sigma}^{\phantom\dagger}\}\right),
\label{Dissipator for ferromagnetic leads}
\end{equation}
where $\ell = A,B$ and $m = +,-$ correspond to the four leads in Fig.~\ref{Fig1}(a), and $\Gamma$ is the rate at which electrons tunnel from the quantum dots to the leads. We have also introduced the jump operators
\begin{equation}
\mathcal{J}_{\ell m}^{\sigma \sigma'}\hat \rho_t \equiv \hat d_{\ell\sigma}^{\phantom\dagger}\hat \rho_t \hat d_{\ell\sigma'}^\dagger (\hat Q_{\ell m})_{\sigma \sigma'},
\end{equation}
where 
\begin{equation}
\hat Q_{\ell m}=\frac{1}{2}(\mathbf{1}+m\zeta_\ell\mathbf{ k}_{\ell}\cdot\boldsymbol{\hat \sigma}),
\end{equation}
and $\mathbf{ k}_\ell$ is a unit vector describing the polarization of detector system $\ell=A,B$, with detector efficiency $\zeta_\ell$, and $\boldsymbol{\hat \sigma}=(\hat \sigma_x,\hat \sigma_y,\hat \sigma_z)$ contains the Pauli matrices. The jump operator describes the transfer of an electron from dot $\ell$ to lead $\ell m$. For the two detector systems, we denote the polarization vectors as $\mathbf{k}_A=\mathbf{ a}$  and $\mathbf{ k}_B=\mathbf{ b}$, respectively. For perfectly polarized leads, we have $\zeta_A=\zeta_B = 1$, and the spin fully determines the probability for an electron to end up in either of the leads. However, for partially polarized leads with  $0\leq \zeta_{A},\zeta_B< 1$, there is a finite probability $1-\zeta_{A,B}$ that an electron ends up in either of the leads regardless of its spin.

The dissipator for local interactions between the dots and the environment is of the form
\begin{equation}
\mathcal{D}_\mathrm{dec}\hat{\rho}_t = \frac{\Gamma_d}{2} \sum_{\ell \sigma \sigma'} \left(\hat L_{\ell}^{\sigma \sigma'}\hat{\rho}_t\hat L_{\ell}^{\sigma \sigma' \dagger}-\frac{1}{2}\{\hat \rho_t, \hat L_{\ell}^{\sigma \sigma' \dagger}\hat L_{\ell}^{\sigma \sigma'} \}\right),
\label{Dissipator for decoherence}
\end{equation}
where $\Gamma_d$ is the decoherence rate (which for the sake of simplicity is assumed to be the same for both quantum dots) and $\hat L_{\ell}^{\sigma \sigma'}$ is a generic Lindblad jump operator. We will primarily focus on decoherence in terms of depolarization, for which the jump operator reads $\hat L_{\ell}^{\sigma \sigma'} = \hat d^\dagger_{\ell \sigma}\hat d_{\ell \sigma'}$. However, we note that other kinds of decoherence can easily be included, such as pure dephasing described by the jump operator $\hat L_{\ell}^{ \sigma \sigma'} = \hat d^\dagger_{\ell \sigma}\hat d_{\ell \sigma}$ \cite{PhysRevLett.114.176803}.

\section{Currents and correlations}
\label{Noise and correlations}

\begin{figure*}[t]
    \centering
    \includegraphics[width=0.97\textwidth]{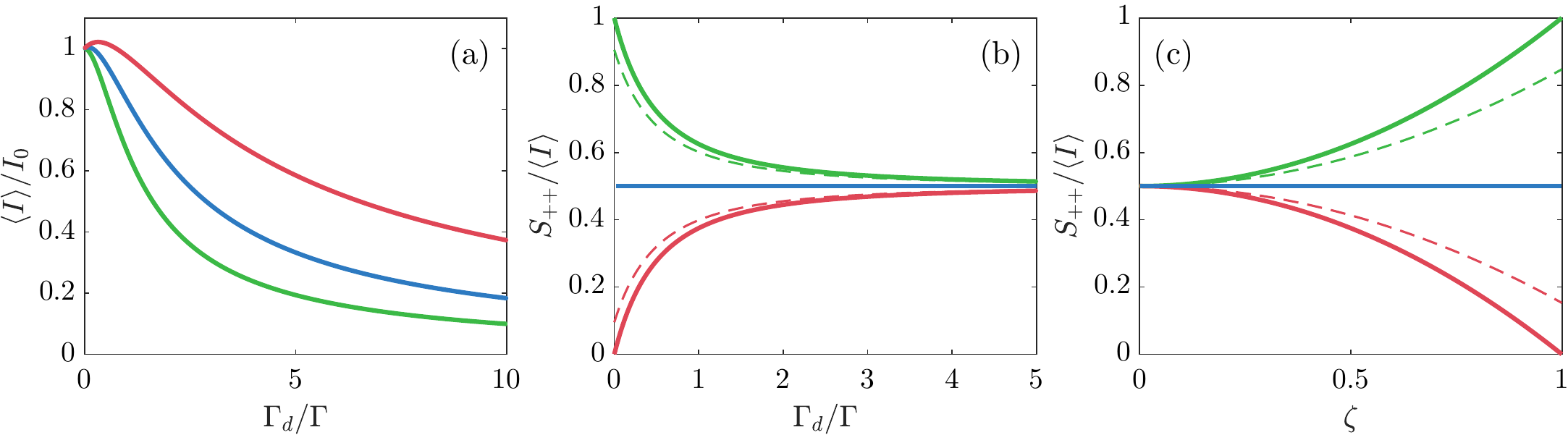}
    \captionsetup{justification=justified,singlelinecheck=false}
        \caption{(a) The average current $\langle I \rangle$, normalized by the current $I_0$ at $\Gamma_d=0$, as a function of the decoherence rate $\Gamma_d$ over the tunnel coupling $\Gamma$ for $\gamma/\Gamma \ll 1$ (green), $\gamma/\Gamma = 0.5$ (blue) and $\gamma/\Gamma = 1$ (red). (b) The cross-correlation $S_{++}$ for anti-parallel (green), orthogonal (blue) and parallel (red) polarizations vectors at $A$ and $B$. The detector effiencies are $\zeta_A=\zeta_B=1$ (solid lines) and $\zeta_A=\zeta_B=0.9$ (dashed lines), respectively. (c) The cross-correlation $S_{++}$ as a function of $\zeta=\zeta_A=\zeta_B$ for anti-parallel (green), orthogonal (blue) and parallel (red) polarization vectors at $A$ and $B$. The decoherence rate is $\Gamma_d/\Gamma=1$ (solid lines) and $\Gamma_d/\Gamma = 0.2$ (dashed lines), respectively. In both panels (b) and (c), we consider the regime $\gamma \ll \Gamma$ and $|\varepsilon_A-\varepsilon_B|\gg \kappa$.}
    \label{Fig2}
\end{figure*}

To investigate how the entanglement of the split Cooper pairs is manifested in the cross-correlations of the drain currents, we consider the relation between the current cross-correlations and the state of the electrons in the dots. To this end, we use techniques from full counting statistics and decompose the density matrix as
\begin{equation}
\hat{\rho}_t = \sum_{\mathbf{m}} \hat{\rho}_t(\mathbf{m}),
\end{equation}
so that $P(\mathbf{m},t) = \text{tr}\left\{ \hat{\rho}_t(\mathbf{m}) \right\}$ is the joint probability that  $\mathbf{m} = (m_{A+},m_{A-},m_{B+},m_{B-})$ electrons have been collected in each lead during the time span $[0,t]$ \cite{Plenio1998,Makhlin2001}. From the Lindblad equation [Eq.~\eqref{Lindblad equation}], we then obtain a hierarchy of coupled equations for $\hat\rho_t(\mathbf{m})$. The equations are decoupled by introducing counting fields $\boldsymbol{\chi} = (\chi_{A+},\chi_{A-},\chi_{B+},\chi_{B-})$ via the transformation
\begin{equation}
\hat\rho_t(\boldsymbol{\chi}) = \sum_{\mathbf{m}}\hat{\rho}_t(\mathbf{m})e^{i\mathbf{m}\cdot \boldsymbol{\chi}}.
\end{equation}
In this way, we obtain a generalized master equation for $\hat\rho_t(\boldsymbol{\chi})$ with a $\chi$-dependent Liouvillian $\mathcal{L}(\boldsymbol{\chi})$, obtained by substituting $\mathcal{J}_{\ell m }^{\sigma \sigma'}\rightarrow e^{i\chi_{\ell m}}\mathcal{J}_{\ell m }^{\sigma \sigma'}$ in Eq.~\eqref{Dissipator for ferromagnetic leads} \cite{Walldorf2020}.

The moment generating function for the number of electrons transferred from the superconductor to the leads during the time span $[0,t]$ now reads
\begin{equation}
    M(\boldsymbol{\chi},t) = \text{tr}\!\left\{\hat\rho_t(\boldsymbol{\chi}) \right\}=\text{tr}\!\left\{ e^{\mathcal{L}(\boldsymbol{\chi})t}\hat\rho_\mathrm{st} \right\},
    \label{MGF}
\end{equation}
where $\hat\rho_\mathrm{st}$ is the stationary state fulfilling $\mathcal{L}(0)\hat{\rho}_\mathrm{st} =0$. The cumulants of the currents are then given by derivatives of the scaled cumulant generating function with respect to the counting fields, evaluated at $\boldsymbol{\chi}=0$,
\begin{equation}
    \langle \!\langle I_{\ell m}^k I_{\ell'm'}^l\rangle\!\rangle = \partial_{i \chi_{\ell m}}^k\partial_{i \chi_{\ell'm'}}^l F(\boldsymbol{\chi})\big|_{\boldsymbol{\chi}=0},
\end{equation}
where the  scaled cumulant generating function, 
\begin{equation}
F(\boldsymbol{\chi})=\lim_{t\rightarrow \infty}\ln[M(\boldsymbol{\chi},t)]/t=\max_j\{\lambda_j(\chi)\},
\end{equation}
is given by the eigenvalue of $\mathcal{L}(\boldsymbol{\chi})$ with the largest real part.\cite{Bagrets2003,Flindt2005}
For instance, the first cumulant is the average current in lead $\ell m$ and it reads (here with $e=1$)
\begin{equation}
   \langle I \rangle \equiv \langle\!\langle I_{\ell m}\rangle\!\rangle = \partial_{i \chi_{\ell m}} F(\boldsymbol{\chi})|_{\boldsymbol{\chi}=0}.
\end{equation}
The average current is shown in Fig.~\ref{Fig2}(a) as a function of the decoherence rate $\Gamma_d$, for three different values of the amplitude of Cooper pair splitting over the coupling to the leads, $\gamma/\Gamma$. As seen in the figure, the current decreases as the decoherence rate increases. The decrease is more pronounced for smaller ratios of $\gamma/\Gamma$. Importantly, the average current is independent of the detector efficiencies $\zeta_A$ and $\zeta_B$ and the polarization vectors $\mathbf{a}$  and $\mathbf{b}$. For $\gamma \ll \Gamma$, we find the simple expression
\begin{equation}
\langle I \rangle = \frac{\gamma^2}{\Gamma+\Gamma_d}.
\end{equation}

The current cross-correlations between lead $A\pm$ and lead $B\pm$ are obtained as the second derivatives
\begin{equation}
S_{\pm\pm} \equiv \langle \!\langle I_{A\pm } I_{B\pm }\rangle\!\rangle = \partial_{i \chi_{A\pm }}\partial_{i \chi_{B\pm }} F(\boldsymbol{\chi})\big|_{\boldsymbol{\chi}=0}.
\end{equation}
Focusing on the regime $\gamma, \Gamma_d \ll \Gamma$, where the emissions of Cooper pairs are well-separated and uncorrelated, we find that the cross-correlations can be expressed as
\begin{equation}
     S_{\pm\pm} = \frac{\langle I \rangle}{2}\text{tr}\left\{(\boldsymbol{1}\pm\zeta_{A}\textbf{a}\cdot\boldsymbol{\hat \sigma})\otimes(\boldsymbol{1}\pm\zeta_{B}\textbf{b}\cdot\boldsymbol{\hat \sigma})\hat\rho_\mathrm{st}^{(2)}\right\},
      \label{Single current cross-correlation}
\end{equation}
which is consistent with earlier works.\cite{Malkoc2014,Brange2017} Here, we have projected the stationary state onto the two-particle sector, which is given simply by the singlet state $\hat \rho_\mathrm{st}^{(2)} =\hat \rho_S= \hat d_S^\dagger|0\rangle\langle 0|\hat d_S$. Importantly, we see that, in this limit, a correlation measurement gives direct access to the statistical properties of the individual pairs, and by changing the polarization vectors $\textbf{a}$ and $\textbf{b}$, one can probe the correlations of the two-particle state. By contrast, if the rate of Cooper pair splitting is on the order of the coupling to the drains, $\gamma\simeq \Gamma$, the Coulomb interactions will introduce correlations between the split Cooper pairs, and a correlation measurement will no longer only reflect the correlations within each split Cooper pair.

A similar situation arises, if the decoherence rate is on the order of the coupling to the drains, $\Gamma_d\simeq \Gamma$. In that case, the cross-correlations in Eq.~\eqref{Single current cross-correlation} cannot be expressed only in terms of the two-particle sector of the stationary state. Specifically, there will be additional correlations that arise after the first electron has left the quantum dots, and the spin has been projected along the corresponding polarization axis. The spin of the remaining electron will be projected into the opposite direction. However, before this electron tunnels into the drains, its spin may undergo further decoherence, which influences the measured correlations. Still, it turns out that the correlations again can be written as in Eq.~\eqref{Single current cross-correlation} provided that the stationary state is replaced by the state
\begin{equation}
   \hat \rho^{(2)} =  \int_0^\infty\!\! dt_1  p(\Gamma,t_1) e^{\mathcal{D}^{(1)}_\mathrm{dec}\otimes\mathbf{1} t_1} \int_0^\infty\!\! dt_2 p(2\Gamma,t_2) e^{\mathcal{D}_\mathrm{dec}t_2} \hat \rho_S,
   \label{Probed state}
\end{equation}
which takes into account how the singlet state decoheres in two steps. In this expression, $p(\Gamma,t) = \Gamma e^{-\Gamma t}$ is the distribution of the time it takes an electron to leave the quantum dots, and $\mathcal{D}^{(1)}_\mathrm{dec}$ is the dissipator in Eq.~\eqref{Dissipator for decoherence} acting only on one of the particles. The integral over $t_2$ describes the average over the time-evolution of the system, while both particles are still in the quantum dots. On the other hand, the integral over $t_1$ describes the average over the time-evolution, when there is only one particle left in the quantum dots. For local dissipators, which cannot produce entanglement, the decohered state is always less entangled than the initial two-particle state injected by the superconductor (here, the singlet state). Thus, while the entanglement witness formulated below is based on expectation values with respect to the decohered state in Eq.~\eqref{Probed state}, any signature of the witness signaling entanglement in the decohered state also indicates that the initial state is entangled. In turn, with a large decoherence rate, the state in Eq.~\eqref{Probed state} may not be entangled, even if the emitted electrons in fact are entangled.

In Fig.~\ref{Fig2}(b), the current cross-correlation $S_{++}$ is shown as a function of the decoherence rate $\Gamma_d$ for different choices of the polarization vectors. While orthogonal measurements are insensitive to decoherence, parallel and anti-parallel measurements converge to the uncorrelated value $S_{++}/\langle I\rangle=1/2$ as the decoherence rate increases. Non-ideal detector efficiencies have a similar effect as shown in Fig.~\ref{Fig2}(c). In other words, both decoherence and imperfect detectors lead to a reduction of the measured correlations.

\section{Entanglement witness}
\label{Entanglement witness}

We are now ready to formulate our entanglement witness based on the correlators in Eq.~\eqref{Single current cross-correlation}. To this end, we introduce an operator representing $N$ current cross-correlation measurements (up to a constant $\langle I \rangle/2$)\cite{Malkoc2014,Brange2017}
\begin{equation}
     \hat W^{(N)} = \sum_{i=1}^{N}(\boldsymbol{1}+\zeta_{A}\textbf{a}_i\cdot\boldsymbol{\hat\sigma})\otimes(\boldsymbol{1}+\zeta_{B}\textbf{b}_i\cdot\boldsymbol{\hat\sigma}),
      \label{CCC operator}
\end{equation}
where $\textbf{a}_i$ and $\textbf{b}_i$ are the polarization vectors of the $i$'th measurement setting, and we recall that $\zeta_A$ and $\zeta_B$ are the detector efficiencies. To ensure that the measurements are as simple as possible, we only consider cross-correlations between one pair of leads, in this case, $A+$ and $B+$, as illustrated in Fig.~\ref{Fig1}(a).

Our main aim for the rest of the paper is now to investigate under what conditions the expectation value of $\hat W^{(N)}$ can be used to detect the entanglement that is generated by the Cooper pair splitter. Specifically, the witness $\hat W^{(N)}$ needs to yield an expectation value for the singlet state or, in general, any state $\hat \rho_e$ whose entanglement we wish to detect, that is different from the expectation values that can be obtained for all separable states $\hat \rho_\mathrm{sep}$. As the set of separable states is convex, this condition implies that the expectation value of the entangled state has to be either larger or smaller than all the ones that can be obtained with the separable states. Without loss of generality, we consider the upper bound condition,
\begin{equation}
    \text{tr}\left\{\hat{W}^{(N)} \hat \rho_e \right\} > \max_{\rho_\mathrm{sep}}\text{tr}\left\{\hat{W}^{(N)}\hat\rho_\mathrm{sep} \right\}.
    \label{Upper bound condition}
\end{equation}
To keep the results as general as possible and independent of our specific model of a Cooper pair splitter, we include all possible separable states in the maximization carried out in Eq.~\eqref{Upper bound condition}, even those that may not be directly relevant for our model. Furthermore, we define the detection margin as the difference between the expectation value of the entangled state we want to detect and the closest expectation value that any separable state may yield,
\begin{equation}
    \Delta \equiv \text{tr}\left\{\hat{W}^{(N)} \hat \rho_e \right\} - \max_{\rho_\mathrm{sep}}\text{tr}\left\{\hat{W}^{(N)}\hat\rho_\mathrm{sep} \right\}.
    \label{Detection margin}
\end{equation}
The entanglement of $\hat \rho_e$ is detectable by $\hat W^{(N)}$ whenever the detection margin is strictly positive, $\Delta > 0$.

Aiming at making the entanglement detection as experimentally feasible as possible, we wish to minimize the number of measurements settings $\mathbf{a}_i$ and $\mathbf{b}_i$ needed to detect the entanglement. For  only one measurement setting, $N=1$, $\hat W^{(1)}$ is a tensor product of local operators, and such an operator cannot detect any entanglement as the largest (smallest) expectation value can always be produced by a separable state. By contrast, for $N=2$, earlier work\cite{Brange2017} has shown that $\hat W^{(2)}$, for certain choices of detector efficiencies and polarization vectors, can detect any entangled pure state, except the maximally entangled. Since the expected state to be produced in a Cooper pair splitter -- the singlet state -- is maximally entangled, we consider $N=3$ measurement settings in the following and for the sake of brevity we set $\hat W \equiv \hat W^{(3)}$.

\section{Witnessing the singlet state}
\label{Detection of the singlet state}

\begin{figure*}[t]
    \centering
    \includegraphics[width=0.97\textwidth]{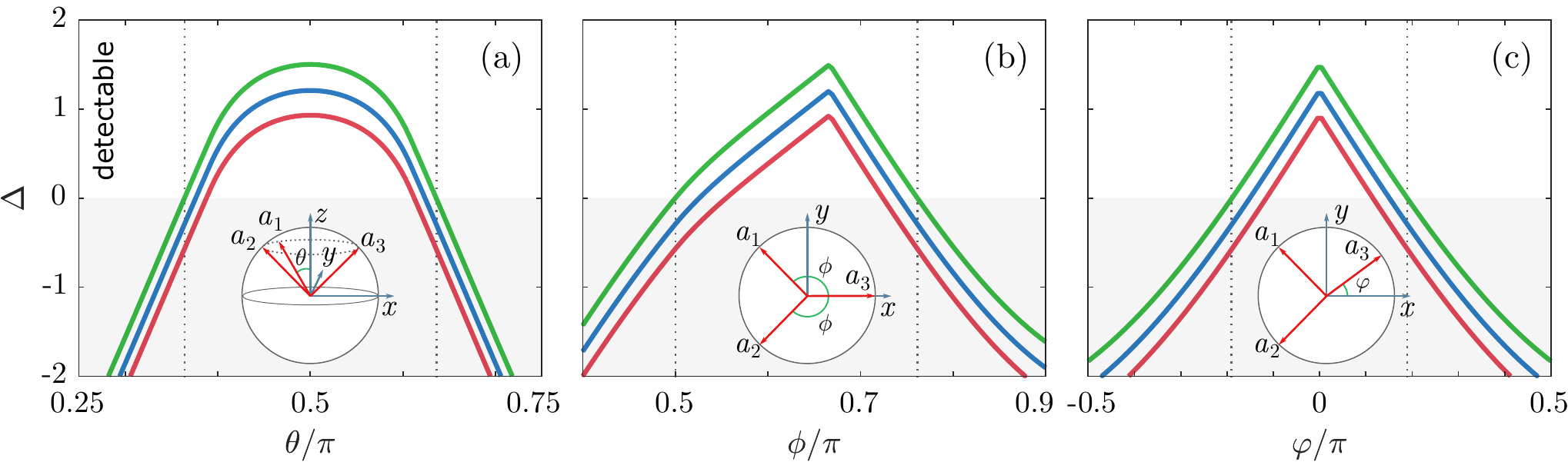}
    \captionsetup{justification=justified,singlelinecheck=false}
        \caption{The detection margin for the singlet state for $\zeta\equiv\zeta_A=\zeta_B = 1$ (green), $\zeta=0.95$ (blue) and $\zeta=0.9$ (red), with known detector efficiencies. (a) As a function of $\theta$, with $\phi = 2\pi/3$ and $\varphi=0$. (b) As a function of $\phi$, with $\theta= \pi/2$ and $\varphi=0$. (c) As a function of $\varphi$, with $\theta=\pi/2$ and $\phi = 2\pi/3$. The insets show the positioning of the polarization vectors at $A$ for each setting.}
    \label{Fig3}
\end{figure*}

Without decoherence, each pair of electrons injected into the quantum dot system is in a maximally entangled singlet state. To find the conditions under which $\hat W$ can detect the entanglement of the singlet state together with the optimal settings that maximize $\Delta$, we consider the explicit expressions for the expectation values of the singlet state and all the separable states in Eq.~\eqref{Upper bound condition}. To this end, we introduce a spherical coordinate system, with angles $\theta_A$, $\phi_A$ and $\varphi_A$, such that the polarization vectors read
\begin{equation}
\begin{split}
    \mathbf{a}_1 &= (\cos[\phi_A]\sin[\theta_A],\sin[\phi_A]\sin[\theta_A],\cos[\theta_A]), \\
    \mathbf{a}_2 &= (\cos[\phi_A]\sin[\theta_A],-\sin[\phi_A]\sin[\theta_A],\cos[\theta_A]), \\
    \mathbf{a}_3 &= \left(\cos[\varphi_A]\sin[\theta_A],\sin[\varphi_A]\sin[\theta_A],\cos[\theta_A]\right),
    \end{split}
        \label{Coordinate angles}
\end{equation}
for detector system $A$, with similar expressions for detector system $B$ in terms of the angles $\theta_B$, $\phi_B$ and $\varphi_B$. Here, we have defined the coordinate system so that the first two polarization vectors have symmetric projections on the $xy$-plane as indicated in the insets in Fig.~\ref{Fig3}.

As shown below, the optimal setting is given by the polarization vectors of the two detector systems being anti-parallel, and radially symmetrically positioned in a plane for each detector system. The difference in Eq.~\eqref{Detection margin} between the expectation value of the singlet state and the largest expectation value of a separable state is then maximized. Depending on whether or not the detector efficiencies of a device are known, the optimization over the expectation values of the separable states has to be done either only for those particular values of the detector efficiencies, or for all detector efficiencies. For known detector efficiencies, we find that the detection margin is
\begin{equation}
    \Delta = \frac{3}{2} \zeta_A\zeta_B,
    \label{Max detection margin for singlet state}
\end{equation}
which is strictly positive whenever $\zeta_A,\zeta_B > 0$. By contrast, for unknown detector efficiencies, the detection margin decreases and becomes
\begin{equation}
    \Delta = 3 \left(\zeta_A\zeta_B-1/2\right),
    \label{Max detection margin for singlet state unknown det eff}
\end{equation}
which means that the entanglement of the singlet state is detectable only if $\zeta_A\zeta_B > 1/2$.

Below, we derive these central results of our work by maximizing the difference between the expectation value for the singlet state and the largest expectation value for the separable states.

\subsection{Expectation value of the singlet state}
In terms of the angles in Eq.~\eqref{Coordinate angles}, the expectation value of $\hat W$ with respect to the singlet state reads
\begin{eqnarray}
 \label{Exp for singlet state}
 \langle \hat W\rangle_e &=& 3-\zeta_A\zeta_B\Big(3\cos\theta_A\cos \theta_B\\
 \nonumber
 &+& \sin\theta_A\sin\theta_B\left[2\cos\left(\phi_A-\phi_B\right)+\cos\left(\varphi_A-\varphi_B\right) \right]\Big).
\end{eqnarray}
This expectation value is maximized when
\begin{equation}
    \theta_A=\pi-\theta_B \equiv \theta, \quad \phi_A=\phi_B\pm \pi\equiv \phi, \quad \varphi_A=\varphi_B\pm \pi\equiv \varphi,
\label{Optimal settings for singlet}
\end{equation}
for which it takes on the maximum value
\begin{equation}
\max_{\{ \mathbf{a}_i\},\{\mathbf{b}_i \}} \langle \hat W\rangle_e = 3\left(1+\zeta_A\zeta_B\right).
\label{Maximal exp value for singlet}
\end{equation}
We note that the condition in Eq.~\eqref{Optimal settings for singlet} means that the expectation value with respect to the singlet state is maximized whenever the polarization vectors of the two detector systems point in opposite directions. This is expected as the singlet state by its nature yields maximal correlations for measurements carried out along opposite directions. Furthermore, we note that the singlet state is an eigenstate of $\hat W$ for the settings in Eq.~\eqref{Optimal settings for singlet} with the expression in Eq.~\eqref{Maximal exp value for singlet} as its eigenvalue.

\subsection{Optimization over the separable states}
We note that the maximal expectation value of $\hat W$ over the convex set of separable states $\hat \rho_\mathrm{sep}$ is obtained for a pure separable state $\hat \rho_\mathrm{sep} = |\Psi_\mathrm{sep}\rangle\langle \Psi_\mathrm{sep}|$; any mixed state only yields a weighted average over pure states. It thus suffices to maximize only over the pure separable states,
\begin{equation}
    \max_{\hat \rho_\mathrm{sep}} \text{tr}\{\hat W \hat \rho_\mathrm{sep}\} = \max_{|\Psi_\mathrm{sep}\rangle} \langle \Psi_\mathrm{sep}|\hat W |\Psi_\mathrm{sep}\rangle.
\end{equation}
Every pure separable state may be represented by two unit Bloch vectors $\textbf{n}_A$ and $\textbf{n}_B$ that describe the local state at each quantum dot. The expectation value of $\hat W$ for a pure separable state then becomes
\begin{equation}
    \langle \hat W \rangle_\mathrm{sep} = \sum_{i=1}^3 \left(1+\zeta_A \mathbf{a}_i\cdot \mathbf{n}_A \right) \left(1+\zeta_B \mathbf{b}_i\cdot \mathbf{n}_B \right).
    \label{Exp value for pure separable state}
\end{equation}
As shown in the Appendix, we find that the detection margin is maximized for the detector settings $\sum_{i}\mathbf{a}_i = \sum_{i}\mathbf{b}_i = 0$ and $\mathbf{a}_i = -\mathbf{b}_i$, for which the largest expectation value over the separable states is
\begin{equation}
    \max_{\hat \rho_\mathrm{sep}} \langle \hat W\rangle = 3\left(1+\zeta_A\zeta_B/2\right).
    \label{Max over sep states}
\end{equation}
In terms of our introduced angles, the optimal settings correspond to $\theta = \pi/2$, $\phi = 2\pi/3$ and $\varphi=0$, or, geometrically, that the polarization vectors are positioned radially symmetric in a plane, and with opposite directions at $A$ and $B$. By subtracting Eq.~\eqref{Max over sep states} from Eq.~\eqref{Maximal exp value for singlet}, we directly obtain the maximal detection margin for a setup with known detector efficiencies [see Eq.~\eqref{Max detection margin for singlet state}]. For unknown detector efficiencies, the measured expectation value has to be compared with the largest expectation value that any separable state, for any detector efficiencies, can yield. Thus, the maximal detection margin in this case [see Eq.~\eqref{Max detection margin for singlet state unknown det eff}] is obtained by subtracting Eq.~\eqref{Max over sep states}, with $\zeta_A=\zeta_B=1$, from Eq.~\eqref{Maximal exp value for singlet}.

To understand how the detection margin changes due to deviations from the optimal settings, we plot it in Fig.~\ref{Fig3} as a function of the angles $\theta$, $\phi$ and $\varphi$. Indeed, we see that the detection margin is maximized for $\theta = \pi/2$, $\phi = 2\pi/3$ and $\varphi = 0$. A deviation from the optimal setting reduces the detection margin, however, the entanglement of the singlet state may still be detected as the detection margin is positive also in a region around these parameter values. Lower detector efficiencies decrease the tolerance for deviations from the optimal settings, showing the importance of having high detector efficiencies, if the polarization vectors cannot be accurately controlled.

\section{Entanglement of pure states}
\label{Detection of less entangled pure states}

\begin{figure*}
    \centering
    \includegraphics[width=0.97\textwidth]{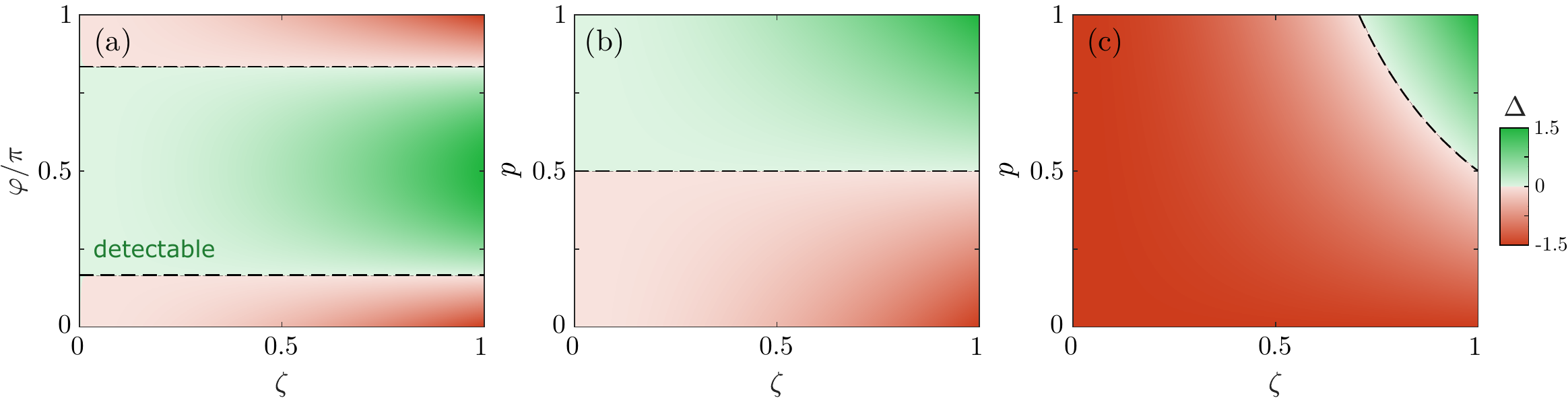}
    \captionsetup{justification=justified,singlelinecheck=false}
    \caption{(a) The detection margin for pure states as in Eq.~\eqref{eq:purestate} with the optimal settings and known detector efficiencies. (b,c) The optimal detection margin $\Delta$ for the Werner state in Eq.~\eqref{Werner state} as a function of $p$ and the detector efficiency $\zeta \equiv \zeta_A=\zeta_B$ for known (b) and  unknown (c) detector efficiencies.}
    \label{Fig4}
\end{figure*}

In addition to detecting the entanglement generated by the Cooper pair splitter, our witness can be used to detect the entanglement of pure states of the form
\begin{equation} 
    \ket{\Psi} = \cos(\varphi/2) \ket{\uparrow}_A \ket{\downarrow}_B - \sin(\varphi/2)\ket{\downarrow}_A\ket{\uparrow}_B
    \label{eq:purestate}
\end{equation}
with the optimal settings for the singlet state and with known detector efficiencies. Here, the angle $0\leq \varphi \leq \pi$ determines the degree of entanglement of the pure state; for $\varphi = 0,\pi$ the state is separable, while it is maximally entangled for $\varphi = \pi/2$. In this case, the analytic expression for the detection margin reads
\begin{equation}
    \Delta = 3 \zeta_A \zeta_B\left[\sin(\varphi)-1/2\right],
    \label{Det marginal for pure states}
\end{equation}
which is strictly positive for $\pi/6 < \varphi/\pi < 5\pi/6$. It is thus possible to detect the entanglement of many other pure states than just the singlet state, however, not of all of them. Furthermore, we see that the detection margin is maximized for the singlet state. A value close to $\Delta \simeq 3/2$ thus indicates not only the presence of an entangled state in the Cooper pair splitter, but also that the state is highly entangled. In addition, the detection margin is reduced, if the detector efficiencies are not known. In that case, the detection margin reads
\begin{equation}
    \Delta = 3 \left[\zeta_A \zeta_B \sin(\varphi)-1/2\right].
    \label{Det marginal for pure states}
\end{equation}
For a symmetric setup with $\zeta\equiv \zeta_A=\zeta_B$, it is then possible to detect the entanglement of pure states with $\sin(\varphi) > 1/(2\zeta^2)$. In Fig.~\ref{Fig4}(a), we show the detection margin for pure states and known detector efficiencies.

\section{Entanglement of mixed states}
\label{Detection of decohered states}
Coming back to our Cooper pair splitter, we now focus on the influence of decoherence on the detection of entanglement. With a finite decoherence rate, the singlet state degrades into a less entangled state, and the entanglement may eventually be lost altogether. In the presence of depolarization [see Eq.~\eqref{Dissipator for decoherence}], we find from Eq.~\eqref{Probed state} that the two-particle state probed by the current cross-correlations is a Werner state\cite{PhysRevA.40.4277}
\begin{equation}
    \hat \rho = p \hat \rho_S + (1-p)\mathbf{1}/4,
    \label{Werner state}
\end{equation}
where $p = 1/(1+\Gamma_d/\Gamma)^2$ determines the weights of the singlet state $\hat \rho_S$ and the maximally mixed state $\mathbf{1}/4$. The Werner state is entangled for $p>1/3$.

Importantly, the expectation value with respect to the second term in Eq.~\eqref{Werner state} does not depend on the detector settings. The expectation value for the Werner state is thus maximized for the same detector settings as the singlet state. Specifically, we find that is becomes
\begin{equation}
    \langle \hat W\rangle_e = 3\left(1+\zeta_A\zeta_B p\right),
\end{equation}
which generalizes the result for the singlet state in Eq.~\eqref{Maximal exp value for singlet}. For known detector efficiencies, the resulting optimal detection margin becomes
\begin{equation}
    \Delta = 3\zeta_A\zeta_B(p-1/2).
    \label{Det marginal for decohered singlet state}
\end{equation}
In Fig.~\ref{Fig4}(b), the detection margin is plotted as a function of $p$ and the (known) detector efficiency $\zeta\equiv \zeta_A=\zeta_B$. For $p=1$ and $\zeta=1$, we indeed recover the maximal detection margin $\Delta =3/2$ for a singlet state. However, for $p < 1$, the detection margin gradually decreases and at $p=1/2$ it becomes negative. Thus, the entanglement of Werner states with $1/3<p \leq 1/2$ is not detectable. The noise tolerance is still significantly better than for entanglement detection with only two current cross-correlation measurements, for which the entanglement cannot be detected for any Werner state with $p\lesssim 0.8$ \cite{Brange2017}.

For unknown detector efficiencies, we instead find
\begin{equation}
    \Delta = 3\left(\zeta_A\zeta_B p-1/2\right),
    \label{Det marginal for decohered singlet state unknown det eff}
\end{equation}
which is plotted in Fig.~\ref{Fig4}(c). We see that the Werner states whose entanglement can be detected with unknown detector efficiencies is much smaller than with known detector efficiencies, and entanglement detection is only possible for $p > 1/(2\zeta_A\zeta_B)$. Moreover, for $\zeta_A\zeta_B<1/2$, entanglement cannot be detected for any value of the mixing parameter, $0\leq p\leq 1$.

\section{Experimental perspectives}
\label{exp_pers}

Finally, we comment on the experimental perspectives of detecting the entanglement generated by a Cooper pair splitter. Based on our analysis, we have found that three current cross-correlation measurements suffice to detect the entanglement of the singlet state. The polarization vectors should be positioned symmetrically in a plane with the angles $\theta = \pi/2$, $\phi=2\pi/3$ and $\varphi=0$. Realistically, the coupling between the quantum dots and the superconductor can be on the order of  $\gamma \simeq1$~MHz, while the coupling to the normal-state electrodes can be around $\Gamma \simeq 100$~MHz, implying that the condition $\gamma\ll \Gamma$ is fulfilled, and the Cooper pairs are emitted well separated and uncorrelated in time. In Fig.~\ref{Fig1}(b), we show the resulting expectation value of the witness operator for three different values of the detector efficiencies, while elastic cotunneling is strongly suppressed by a large detuning of the quantum dot levels. We show the entanglement witness as a function of the decoherence rate over the tunnel coupling, and we see that the entanglement can be detected, if the decoherence rate can be kept below the escape rate to the leads. With unknown detector efficiencies, we find from Eq.~\eqref{Det marginal for decohered singlet state unknown det eff} that the entanglement can be detected  as long as $\Gamma_d < \left(\sqrt{2\zeta_A\zeta_B}-1\right)\Gamma$, which agrees well with the results in Fig.~\ref{Fig1}(b). For known detector efficiencies, the detector margin improves [see Eq.~\eqref{Det marginal for decohered singlet state}], and the entanglement can be detected as long as $\Gamma_d < \left(\sqrt{2}-1\right)\Gamma$. In both cases, the entanglement can be detected even with a rather large decoherence rate.

It is worth also to mention alternative realizations of our entanglement witness. If the ferromagnetic leads do not easily allow for a rotation of the polarization axes, it may instead be possible to rotate the individual spins in the quantum dots by using a material with a strong spin-orbit coupling combined with oscillating electric fields.\cite{Flindt2006,Nowack2007,Hanson2007} Thus, instead of rotating the polarization vectors of the detectors, one could rotate the spins before they are detected. It may also be possible to perform real-time detection of the electrons in the quantum dots as in the experiment of Ref.~\onlinecite{Ranni2021} and perform others types of spin-to-charge read-out of the spins,\cite{Hanson2007} instead of using ferromagnetic leads. It should also be noted that our entanglement witness is not restricted to static Cooper pair splitters. It can also be applied to the dynamic Cooper pair splitter described in Ref.~\onlinecite{Brange2021}.

\section{Conclusions and outlook}
\label{Conclusion and outlook}
We have formulated an entanglement witness that can detect the entanglement generated by Cooper pair splitters using only three current cross-correlation measurements, and we have determined the maximum rate of decoherence for the entanglement to be detectable. The small number of correlation measurements is promising for an experimental implementation of our witness as opposed to conventional Bell inequalities, which typically require many more cross-correlation measurements. We have also found that the optimal detector setting is to position the three polarization vectors of each detector system radially symmetrically in a plane. For such a setting, the polarization vectors are non-collinear, a necessary condition to detect entanglement. This is an important difference to witnesses based on only two correlation measurements, for which the corresponding symmetric setting yields collinear polarization vectors (which is thus incapable of detecting the entanglement of the maximally entangled states). Furthermore, for spin read-out with ferromagnetic leads, the positioning of the polarization vectors in a plane means that it is not necessary to rotate the magnetic polarization in all three dimensions. It is possible that such a symmetry is also optimal for detection schemes with a larger number of cross-correlation measurements.

Our entanglement witness provides a feasible way of detecting the entanglement produced by  Cooper pair splitters. Given the recent progress in controlling and detecting individual electrons in such devices \cite{Ranni2021}, our findings may pave the way for an experiment, where the spin entanglement between mobile electrons is detected.

\acknowledgements
The work was supported by Academy of Finland through the Finnish Centre of Excellence in Quantum Technology (project numbers 312057 and 312299) and grants number 308515 and 331737.

\newpage
\onecolumngrid

\appendix
\section*{Appendix: Maximization of the detection margin for the singlet state}
\label{Appendix}
\addcontentsline{toc}{section}{Appendices}

Here we derive the optimal detector settings (in terms of the angles $\theta_A,\phi_A,\varphi_A,\theta_B,\phi_B,\varphi_B$) which maximize the detection margin for the singlet state, i.e., the difference between the expectation value of the singlet state and the largest one of the separable states. To this end, we first consider the expectation value of a pure separable state
\begin{equation}
    \langle \hat W \rangle_\mathrm{sep} = \sum_{i=1}^3 \left(1+\zeta_A \mathbf{a}_i\cdot \mathbf{n}_A \right) \left(1+\zeta_B \mathbf{b}_i\cdot \mathbf{n}_B \right),
    \label{Exp sep states appendix}
\end{equation}
where $\mathbf{n}_A$ and $\mathbf{n}_B$ are Bloch vectors defining the separable state at each quantum dot. The maximum value for $\mathbf{n}_B$ is obtained when
\begin{equation}
\mathbf{n}_B = \frac{\sum_{i=1}^3  \left(1+\zeta_A \mathbf{a}_i\cdot \mathbf{n}_A \right)\mathbf{b}_i}{\left\| \sum_{i=1}^3  \left(1+\zeta_A \mathbf{a}_i\cdot \mathbf{n}_A \right)\mathbf{b}_i\right\|},
\end{equation}
for which we find
\begin{equation}
    \max_{\mathbf{n}_B} \langle \hat W \rangle_\mathrm{sep} = 3+\zeta_A \sum_{i=1}^3\mathbf{a}_i\cdot \mathbf{n}_A+ \zeta_B \left\|\sum_{i=1}^3(1+\zeta_A\mathbf{a}_i\cdot \mathbf{n}_A) \mathbf{b}_i\right\|,
\end{equation}
where $\left\|\mathbf{u}\right\|=\sqrt{u_x^2+u_y^2+u_z^2}$ denotes the norm of a vector $\mathbf{u}=(u_x,u_y,u_z)$. For any $\mathbf{n}_A$ and $\{\mathbf{a}_i\}$, the norm is minimized, if $\{\mathbf{b}_i\}$ lies in a plane. To minimize the largest expectation value of the separable states compared to the singlet state, we thus need to set $\theta_B=\pi/2$. Furthermore, applying the same arguments with $A$ and $B$ swapped, we find
\begin{equation}
    \max_{\mathbf{n}_A,\mathbf{n}_B} \langle\hat W\rangle_\mathrm{sep}(\theta_A,\theta_B) \geq \max_{\mathbf{n}_A,\mathbf{n}_B} \langle \hat W\rangle_\mathrm{sep}(\theta_A,\pi/2)\geq \max_{\mathbf{n}_A,\mathbf{n}_B} \langle \hat W\rangle_\mathrm{sep}(\pi/2,\pi/2),
\end{equation}
where we, for the sake of brevity, have left out the explicit dependence on the other angles. We conclude that $\theta_A=\theta_B=\pi/2$ minimizes the largest expectation value of the separable states for any given values of $\phi_A, \varphi_A, \phi_B, \varphi_B$. Next, to determine the optimal values of the remaining angles, we consider Eq.~\eqref{Exp sep states appendix}, which we now express as
\begin{eqnarray}
 \langle \hat W \rangle_\mathrm{sep}(\theta_A=\theta_B=\pi/2) &=& 3+ \zeta_A D(\phi_A,\varphi_A,\phi_1)\sin[\theta_1]+\zeta_BD(\phi_B,\varphi_B,\phi_2)\sin[\theta_2] \\
\nonumber
 &+&\frac{\zeta_A\zeta_B}{2}\sin[\theta_1]\sin[\theta_2]\left[ D(\phi_A+\phi_B,\varphi_A+\varphi_B,\phi_1+\phi_2)
    +D(\phi_A-\phi_B,\varphi_A-\varphi_B,\phi_1-\phi_2)\right].
\end{eqnarray}
where we have defined the function
\begin{equation}
   D(x,y,z) \equiv \left[2 \cos(x)+\cos(y)  \right] \cos(z) + \sin(y)\sin(z).
\end{equation}
Here, the two angles $\phi_1$ and $\phi_2$ are the azimuths of the two Bloch vectors $\mathbf{n}_A$ and $\mathbf{n}_B$, while $\theta_1$ and $\theta_2$ are the inclinations, for instance, we have $\mathbf{n}_A = (\cos[\phi_1]\sin[\theta_1],\sin[\phi_1]\sin[\theta_1],\cos[\theta_1])$. We now directly see that to find the largest possible value of $\langle \hat W\rangle_\text{sep}$ with respect to $\theta_1$ and $\theta_2$, we need to set $\theta_1=\theta_2=\pi/2$.

Next, using the ansatz, $\phi_1=\phi_2+\pi$, we find a lower bound for the largest expectation value of the separable states,
\begin{eqnarray}
\nonumber
 &&\min_{\theta_A,\theta_B}\max_{\mathbf{n}_A,\mathbf{n}_B}\langle \hat W \rangle_\mathrm{sep} = \max_{\phi_1,\phi_2}\langle \hat W \rangle_\mathrm{sep}(\theta_{A,B}=\pi/2,\theta_{1,2}=\pi/2) \geq \max_{\phi_1}\langle \hat W \rangle_\mathrm{sep}(\theta_{A,B}=\pi/2,\theta_{1,2}=\pi/2,\phi_2=\phi_1-\pi) \\
 \nonumber
 &=& 3+ \zeta_A D(\phi_A,\varphi_A,\phi_1)+\zeta_BD(\phi_B,\varphi_B,\phi_1-\pi) +\frac{\zeta_A\zeta_B}{2}\Big[ D(\phi_A+\phi_B,\varphi_A+\varphi_B,2\phi_1-\pi)
    +D(\phi_A-\phi_B,\varphi_A-\varphi_B,\pi)\Big] \\
    \nonumber
&=& 3+ \sum_{i=1}^3 \left(\zeta_A \mathbf{a}_i-\zeta_B \mathbf{b}_i \right)\cdot \mathbf{n}_A-\frac{\zeta_A\zeta_B}{2}\Big[ D(\phi_A+\phi_B,\varphi_A+\varphi_B,2\phi_1)
    +D(\phi_A-\phi_B,\varphi_A-\varphi_B,0)\Big] \\
    &\geq& 3-\frac{\zeta_A\zeta_B}{2}D(\phi_A-\phi_B,\varphi_A-\varphi_B,0).
\end{eqnarray}
Here, we have used that the second and the third terms in the second last step can always be made non-negative by choosing an appropriate $\phi_1$. Importantly, $D(\phi_A+\phi_B,\varphi_A+\varphi_B,2\phi_1)$ is invariant under a rotation of $\pi$, thus, $\mathbf{n}_A$ can always be chosen (flipped) such that both terms are non-negative. We compare this lower bound for the largest expectation value of the separable states with the expectation value of the singlet state [cf. Eq.~\eqref{Exp for singlet state}]
\begin{equation}
    \langle \hat W\rangle_e = 3-\zeta_A\zeta_BD(\phi_A-\phi_B,\varphi_A-\varphi_B,0).
\end{equation}
We then see that there is a trade-off between having a large expectation value for the singlet state and a small maximal expectation value for the separable states. However, combining the two expressions, we find an upper bound for the detection margin reading
\begin{equation}
    \Delta \leq -\frac{\zeta_A\zeta_B}{2}D(\phi_A-\phi_B,\varphi_A-\varphi_B,0),
\end{equation}
which is maximized for $D(\phi_A-\phi_B,\varphi_A-\varphi_B,0) = -3$. Importantly, the inequality is tight, i.e., $D(\phi_A-\phi_B,\varphi_A-\varphi_B,0) = -3$, or equivalently, $\sum_i \mathbf{a}_i = \sum_i \mathbf{b}_i = 0$ and $\mathbf{a}_i = -\mathbf{b}_i$, yields equality. Since this is the upper bound for the detection margin, it is the optimal setting.

\twocolumngrid

\end{document}